\documentclass[conference]{IEEEtran}
\IEEEoverridecommandlockouts
\usepackage{xcolor} 
\usepackage[noend]{algpseudocode}
\usepackage[numbers,sort&compress]{natbib}
\usepackage{array}
\usepackage{amsfonts}
\usepackage{graphicx}
\usepackage{bm,mathrsfs}
\usepackage{epstopdf}
\usepackage{mathtools}
\usepackage{tikz,pgfplots}
\usetikzlibrary{decorations,arrows,shapes,trees}
\usepackage{amsmath,amsthm}
\usepackage{amsopn}
\usepackage{listings}
\usepackage{adjustbox}
\usepackage{longtable}
\usepackage{multirow}
\usepackage{hyperref}
\usepackage{pgfplots}
\usepackage{pgfplotstable}
\usepackage{grffile}
\usetikzlibrary{arrows,shapes,plotmarks}
\usepgfplotslibrary{groupplots}
\usetikzlibrary{matrix}

\usepackage[font=small,belowskip=-3pt,aboveskip=3pt]{caption}
\newcommand{\cref}[2]{\hyperref[#2]{#1~\ref*{#2}}}
\newcommand{\figref}[1]{\hyperref[#1]{Fig.~\ref*{#1}}}
\newcommand{\secref}[1]{\hyperref[#1]{Section~\ref*{#1}}}
\newcommand{\tabref}[1]{\hyperref[#1]{Tab.~\ref*{#1}}}
\newcommand{\eqnref}[1]{\hyperref[#1]{Eq.~(\ref*{#1})}}
\hypersetup{
	colorlinks,
	linkcolor={blue!50!black},
	citecolor={blue!50!black},
	urlcolor={blue!80!black},
	anchorcolor = {blue!80!black},
	filecolor = {blue!80!black},
	menucolor = {blue!80!black},
	runcolor = {blue!80!black}
}

\pgfplotsset{compat=1.8}
\usepackage{soul}
\usepackage{fancyvrb}
\usepackage{rotating}
\usepackage{url}
\usepackage{algorithm}

\usepackage{enumitem}
\usepackage{subcaption}

\newcommand{\mvec}{\textsc{matvec}}

\newcommand{\dkt}{\textsc{Dendro-KT}}

\newcommand{\Dendro}{\textsc{Dendro}}

\newcommand{\In}{\textsc{In}}
\newcommand{\Out}{\textsc{Out}}

\newcommand{\Frontera}{\href{https://frontera-portal.tacc.utexas.edu/}{Frontera}}
\newcommand{\petsc}{\href{https://www.mcs.anl.gov/petsc/}{PETSc}}

\newcommand{\paraview}{\href{https://www.paraview.org/}{ParaView}}
\pgfplotsset{
compat=1.8,
legend image code/.code={
\draw[mark repeat=2,mark phase=2]
plot coordinates {
(0cm,0cm)
(0.15cm,0cm)        
(0.3cm,0cm)         
};%
}
}

\newcommand{\norm}[1]{\left\lVert#1\right\rVert}
\definecolor{cpu3}{HTML}{F44336}
\definecolor{cpu4}{HTML}{2196F3}
\definecolor{cpu1}{HTML}{4CAF50}
\definecolor{cpu2}{HTML}{FFC107}
\definecolor{gpu3}{HTML}{EF9A9A}
\definecolor{gpu4}{HTML}{90CAF9}
\definecolor{gpu1}{HTML}{A5D6A7}
\definecolor{gpu2}{HTML}{FFE082}

\definecolor{cpu5}{HTML}{9932CC}

\definecolor{sq_b1}{RGB}{37,52,148}
\definecolor{sq_b2}{RGB}{44,127,184}
\definecolor{sq_b3}{RGB}{65,182,196}
\definecolor{sq_b4}{RGB}{127,205,187}
\definecolor{sq_b5}{RGB}{199,233,180}
\definecolor{sq_b6}{RGB}{255,255,204}

\definecolor{sq_r1}{RGB}{189,0,38}
\definecolor{sq_r2}{RGB}{240,59,32}
\definecolor{sq_r3}{RGB}{253,141,60}
\definecolor{sq_r4}{RGB}{254,178,76}
\definecolor{sq_r5}{RGB}{254,217,118}
\definecolor{sq_r6}{RGB}{255,255,178}

\definecolor{sq_g1}{RGB}{0,104,55}
\definecolor{sq_g2}{RGB}{49,163,84}
\definecolor{sq_g3}{RGB}{120,198,121}
\definecolor{sq_g4}{RGB}{173,221,142}
\definecolor{sq_g5}{RGB}{217,240,163}
\definecolor{sq_g6}{RGB}{255,255,204}

\definecolor{div_c1}{RGB}{230,171,2}
\definecolor{div_c2}{RGB}{102,166,30}
\definecolor{div_c3}{RGB}{231,41,138}
\definecolor{div_c4}{RGB}{117,112,179}
\definecolor{div_c5}{RGB}{217,95,2}
\definecolor{div_c6}{RGB}{27,158,119}
\definecolor{div_c7}{RGB}{215,48,39}

\definecolor{div_d1}{RGB}{215,25,28}
\definecolor{div_d2}{RGB}{253,174,97}
\definecolor{div_d3}{RGB}{255,255,191}
\definecolor{div_d4}{RGB}{171,217,233}
\definecolor{div_d5}{RGB}{44,123,182}
\definecolor{ao}{RGB}{0.0, 128, 0.0}
\newtheorem*{remark}{Remark}
\usepackage{dirtytalk}

\newcommand{\added}[1]{\textcolor{black}{#1}}
\usepackage[normalem]{ulem}
\usepackage[noadjust]{cite}

\usetikzlibrary{external}
\def\BibTeX{{\rm B\kern-.05em{\sc i\kern-.025em b}\kern-.08em
    T\kern-.1667em\lower.7ex\hbox{E}\kern-.125emX}}
\begin{document}

\title{Case study of SARS-CoV-2 transmission risk assessment in indoor environments using cloud computing resources}

\author{
\IEEEauthorblockN{Kumar Saurabh}
\IEEEauthorblockA{Iowa State University\\
Email: maksbh@iastate.edu}\\   
\IEEEauthorblockN{Masado Ishii}
\IEEEauthorblockA{ University of Utah\\
Email: masado@cs.utah.edu}\\
\IEEEauthorblockN{Hari Sundar}
\IEEEauthorblockA{University of Utah\\
Email: hari@cs.utah.edu}\\
\and
\IEEEauthorblockN{Santi Adavani}
\IEEEauthorblockA{ RocketML Inc.\\
Email: santi@rocketml.net}\\
\IEEEauthorblockN{Boshun Gao }
\IEEEauthorblockA{ Iowa State University\\
Email: boshun@iastate.edu}\\
\and
\IEEEauthorblockN{Kendrick Tan}
\IEEEauthorblockA{Iowa State University\\
Email: ken@iastate.edu}\\
\IEEEauthorblockN{Adarsh Krishnamurthy }
\IEEEauthorblockA{ Iowa State University\\
Email: adarsh@iastate.edu}\\
\IEEEauthorblockN{Baskar Ganapathysubramanian}
\IEEEauthorblockA{Iowa State University\\
Email: baskarg@iastate.edu}
}
\IEEEaftertitletext{\vspace{-3\baselineskip}}
\maketitle
\begin{abstract}
Complex flow simulations are conventionally performed on HPC clusters. However, the limited availability of HPC resources and steep learning curve of executing on traditional supercomputer infrastructure has drawn attention towards deploying flow simulation software on the cloud. We showcase how a complex computational framework--that can evaluate COVID-19 transmission risk in various indoor classroom scenarios--can be abstracted and deployed on cloud services. The availability of such cloud-based personalized planning tools can enable educational institutions, medical institutions, public sector workers (courthouses, police stations, airports, etc.), and other entities to comprehensively evaluate various in-person interaction scenarios for transmission risk. We deploy the simulation framework on the Azure cloud framework, utilizing the ~\dkt{} mesh generation tool and ~\petsc{} solvers. The cloud abstraction is provided by RocketML cloud infrastructure. We compare the performance of the cloud machines with state-of-the-art HPC machine TACC ~\Frontera. Our results suggest that cloud-based HPC resources are a viable strategy for a diverse array of end-users to rapidly and efficiently deploy simulation software.

\end{abstract}


\section{Motivation}
\label{sec:motivation}
Current and next-generation high-performance computing (HPC) resources play a pivotal role in helping answer many of the fundamental questions related to science and technology. However, the limited availability of HPC resources, along with the steep learning curve needed to employ these resources effectively, have limited the utilization of HPC resources to academic and large industrial R\&D groups. Additionally, standard HPC resource utilization policies restrict the number of nodes available to the user and the number of job instances per user. These limitations become a bottleneck, especially in risk assessment and design exploration, where parametric studies need to be performed that involve executing multiple instances of the simulation with different simulation inputs.  In this context, the availability of on-demand cloud services suggests an attractive alternative for adaptive, interactive scientific computing workloads.

The ubiquitous availability of cloud computing suggests a path to the democratization of complex multi-physics simulations. Simulations--for example, parametric studies of ventilation rates and seating arrangements in critical public places such as courthouses, hospital waiting rooms, etc. to evaluate transmission risk--are currently limited to deployment on HPC clusters that require a steep learning curve to deploy, which disenfranchises non-computer savvy scientists, decision-makers, and designers. Additionally, the data visualization and analysis can be done in-situ without copying data from one place to another. In this work, we propose a public cloud-based framework with in-situ visualization and post-processing capabilities for scalable deployment of fluid dynamics simulation using FEM discretization. Our results suggest that the performance of such a framework is at par, if not better, with the framework executed on state-of-the-art HPC resources.

\section{Background}
\label{sec:background}
A large fraction of engineered and natural systems are analyzed using scientific simulation codes that involve numerical solutions of partial differential equations (PDE). The Finite Element Method (FEM) is a widely popular approach for the numerical solution of PDEs in complex domains, with multiple \$ billions/year spent in CAD and CAM (computer-aided design and manufacturing) based FEM software alone~\citep{Hughes_iso_book}. The popularity arises from a compelling set of properties, including the ability (a) to model arbitrary geometries, (b) to change the order of representation (linear, quadratic, and higher-order), (c) to utilize variational arguments that guarantee monotonic convergence to the solution with improved discretization, and (d) to seamlessly utilize {\it a posteriori} error estimates to adapt the mesh. Over the past several years, we have developed a scalable FEM solver based on adaptive octree mesh and deployed this framework for a diverse array of multi-physics simulations--primarily on HPC resources~\citep{khanwale2020fully,saurabh2020industrial,fernando2018massively,xu2021octree,neilsen2018massively,neilsen2019dendro,saurabh2021scalable}.

In this work, we explore the deployment (and assess performance) of this FEM framework on a public cloud computing facility. Cloud computing provides on-demand access to computing resources, data storage, development, and networking facilities via the internet. Cloud computing offers some key benefits compared to the traditional HPC settings, in particular:
\begin{itemize}[left=0pt]
    \item \textbf{Lower IT costs}: Cloud facilities enable the offloading of some or most of the costs and effort of purchasing, installing, configuring, and managing on-premises infrastructure. This can be especially significant for small research groups.
    \item \textbf{Scale on demand}: Instead of purchasing the capacity and waiting for the arrival of hardware, the cloud enables on-demand access and elastic scalability.
    \item \textbf{Choice of architecture}: Cloud allows to choose from different types of architecture, depending on the nature of codes. Some computer architectures are tailored for compute-bound codes, whereas others provide good performance for memory-bound codes. A traditional HPC cluster usually has limited flexibility in terms of the availability of different types of machines. In contrast, cloud computing allows the flexibility to choose the machines based on the nature of the underlying code (compute-bound vs. memory-bound vs. communication-bound).
    \item \textbf{End-to-end deployment}: Cloud enables the full stack of software, starting from development to testing to execution to visualization. The latter aspect--visualization--is particularly appealing, as it allows for in-situ visualization and analysis without extensive data movement. We note that some federal HPC clusters, like TACC, have a dedicated visualization portal but are not accessible on all of them. 
\end{itemize}

In this paper, we seek to address the following three components of the workshop:
\begin{itemize}[left=0pt]
    \item On-demand and interactivity with performance, scaling, and cost efficiencies;
    \item Application environment, integration and deployment technologies;
    \item Workflow orchestration using public cloud and HPC data center resources.
\end{itemize}

\section{Problem Definition}
\label{sec:problemDefinition}
COVID-19 has irreversibly changed how we consider transmission risk in indoor environments. The Six-Foot Rule, for instance, is a guideline that does not account for small aerosol droplets that are continuously mixed through an indoor space. The distribution of aerosolized particles depends on a wide array of factors (ventilation, airflow patterns that are impacted by furniture, as well as the respiratory activity of the inhabitants. While conventional risk assessment tools may indicate that a particular room is low-risk \textit{\textbf{on average}}, there may be specific locations in the room with significantly higher risk for transmission--for example, where there is local recirculation causing limited air exchange with the outside environment. 
Assessing the localized risk becomes especially important if individuals are seated in such locations for extended periods, increasing their cumulative exposure time. We consider scenarios that impact \textit{first-responders, essential workers, and K-12 students}. For instance, courthouse activities require participants (judges, clerks, petitioners, jurors) to remain sedentary over long periods. Similarly, most classroom activities require K-12 students to stay seated for extended periods. In such cases, it becomes imperative to identify if specific locations have a higher risk and rank among alternate seating arrangements. 

In this context, high fidelity fluid/aerosol simulation is a viable approach to evaluate various seating and operating scenarios, identify risk factors, and rank order mitigation strategies. We aim to democratize such simulations on cloud-based resources by integrating four sophisticated concepts to provide very detailed estimates of aerosol distribution in indoor environments: (a) a detailed flow physics simulation (called Large Eddy Simulations) that produces significantly more accurate results while taking reasonable compute time, (b) accounting for the thermal plumes created by heat released by humans and electronics (which changes the flow in rooms), (c) immersogeometric analysis that allows rapid exploration of a diverse array of configurations with complex geometries, (d) a passive scalar transport based model that tracks the spatial distribution of the aerosolized concentration. To our best knowledge, parametric simulations of such high-fidelity simulations have not been deployed on the cloud. The availability of such a cloud-based analysis approach will allow education administrators (K-12), federal/state/local government officials (courthouses, police stations), and hospital administrators to make informed decisions on seating arrangements and operating procedures.

\begin{figure}
    \centering
  \begin{subfigure}{0.48\linewidth}
    \includegraphics[width=0.95\linewidth]{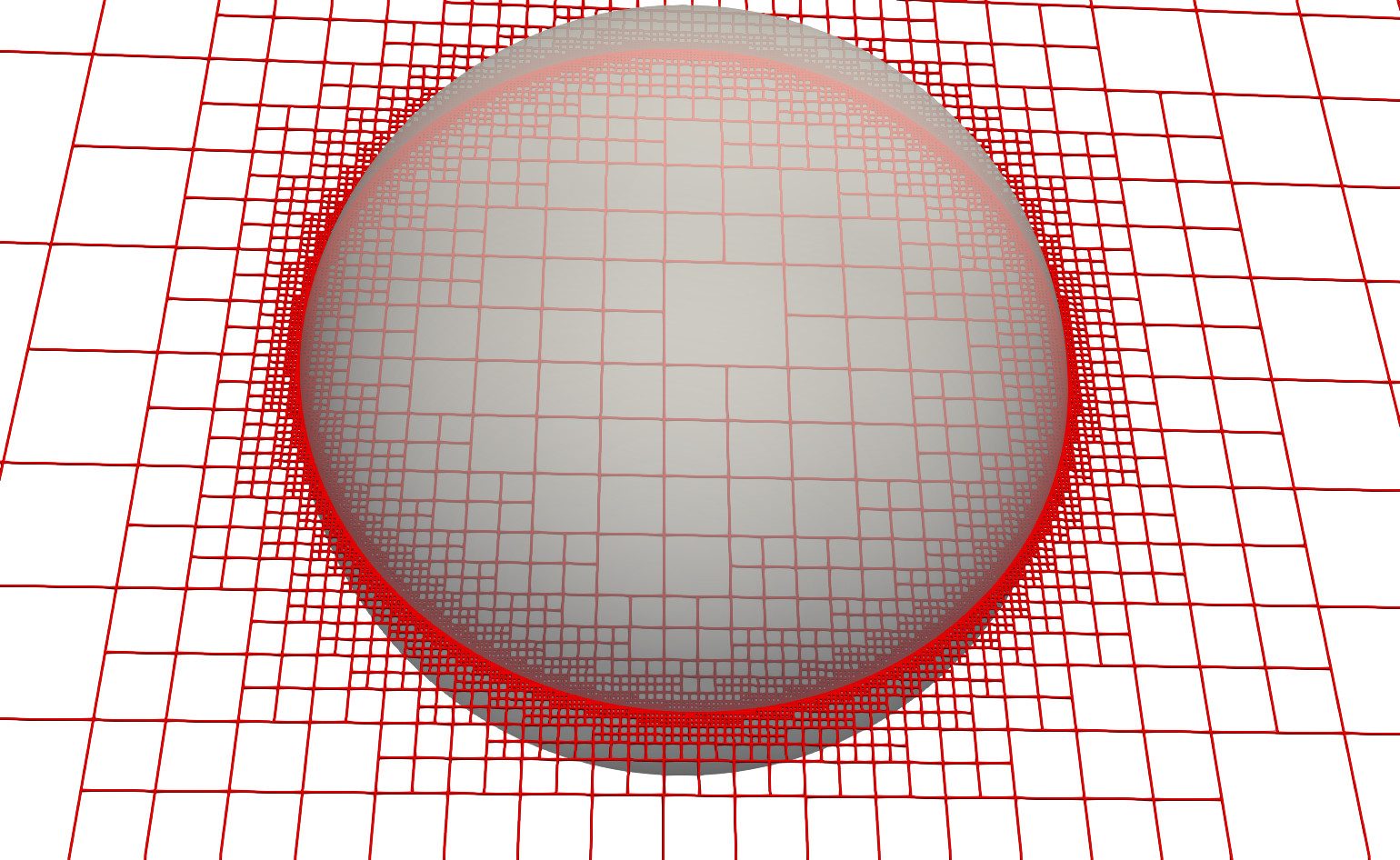}
    \caption{immersed}
    \label{fig:immersed}
\end{subfigure}
\begin{subfigure}{0.48\linewidth}
    \includegraphics[width=0.95\linewidth]{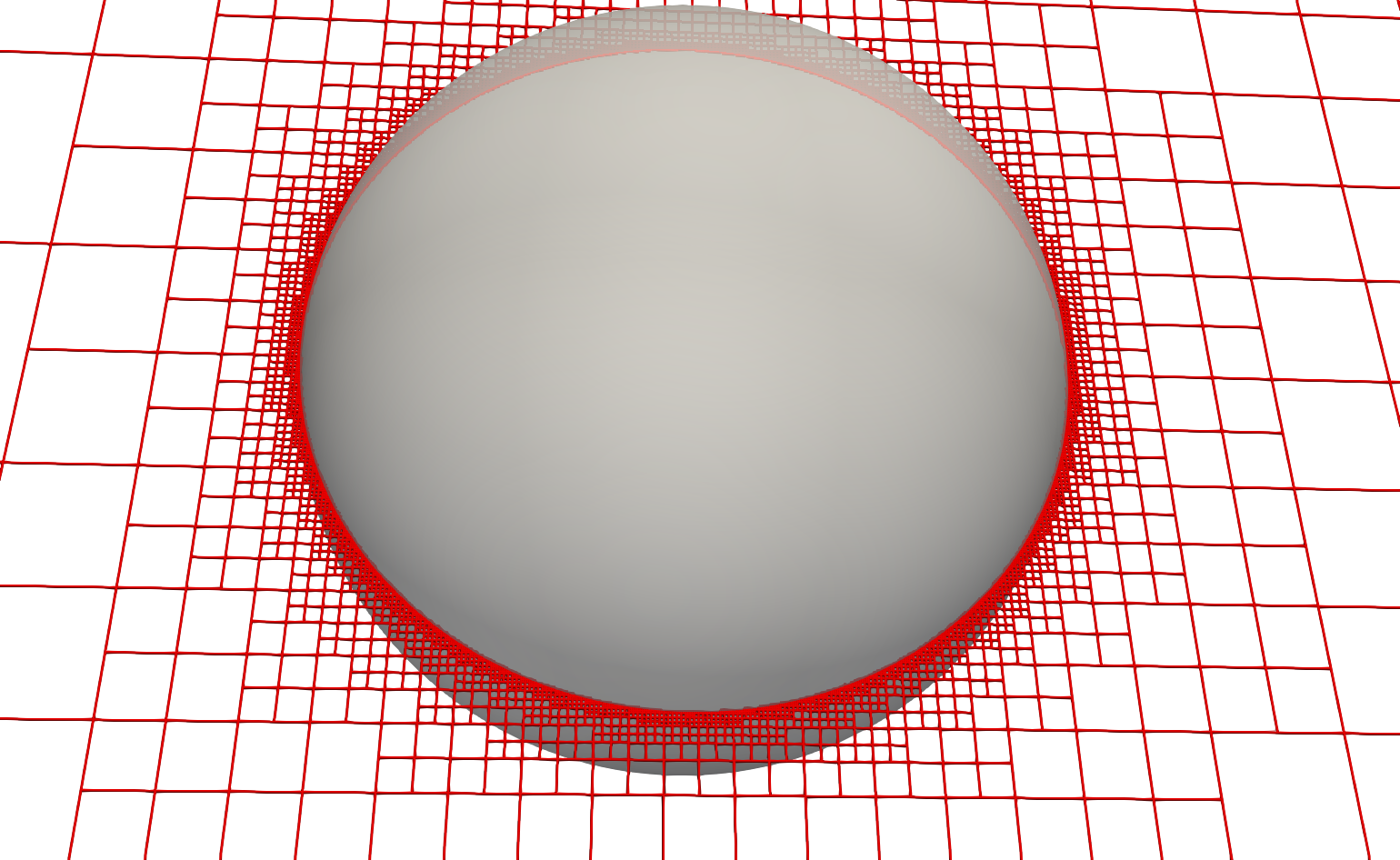}
    \caption{carved out }
    \label{fig:carved}
\end{subfigure}
\vspace{4 mm}
\caption{Difference between the adaptive mesh for \textit{immersed} and \textit{carved} out for the sphere case. In immersed case, we retain the full octree and this gives to a significantly large number of elements and nodes compared to the carved out case. It must be noted the elements that are inside the object do not contribute to the accuracy of the solution. Eventually Dirichlet Boundary condition are imposed on all the \In nodes. \vspace{1 mm}}
\label{fig: meshComparison}
\end{figure}

In this work, we use \dkt{} for efficient and automatic mesh generation. \dkt{} is an extension of previous \Dendro{} versions to incomplete octrees. Tree-based grid generation (quadtrees in 2D and octrees in 3D) is common in computational sciences \citep{tu2005scalable,ishii2019solving,fernando2018massively,sundar2008bottom,bastian2008generic,greaves1999hierarchical,bader2012space,BursteddeWilcoxGhattas11,popinet2003gerris} largely due to its simplicity and parallel scalability. The ability to efficiently refine (and coarsen) regions of interest using tree-based data structures have made it possible to deploy them on large-scale multi-physics simulations~\citep{akhtar2013boiling,saurabh2020industrial,fernando2018massively,khanwale2020fully,rudi2015extreme,bielak2005parallel,losasso2004simulating,kim2003large}. The use of incomplete octrees enables the handling of any arbitrary complex geometries, where the stated geometry is carved out from the domain (\figref{fig: meshComparison}). The framework's simplicity and robustness lie in the fact that complete mesh generation only needs a specification of \In{} or \Out{} (of the geometry) for any queried point. Furthermore, the efficient \mvec{} computation is performed by traversing the trees rather than using any mesh-based data structure to prevent indirect memory access.

Once the mesh is created, we model the time-dependent transmission of the viral load as a scalar transport equation that is advected by a statistically steady-state flow field obtained from the solution of coupled Navier–Stokes and Heat transfer solver.  The well-established Variational Multiscale (VMS) FEM \citep{bazilevs2007variational} is used for the discretization. \dkt{} interfaces with \petsc{} for solving the system of equations. We have recently extensively validated and demonstrated the capabilities of this framework ~\citep{saurabh2020industrial,saurabh2021scalable}.

\section{Workflow}
\label{sec:Workflow}
In this section, we detail and demonstrate the deployment of our framework on the cloud infrastructure. 

\subsection{Cloud Abstraction}

\begin{figure*}[t!]
    \centering
    \includegraphics[width=0.98\textwidth,height=80mm]{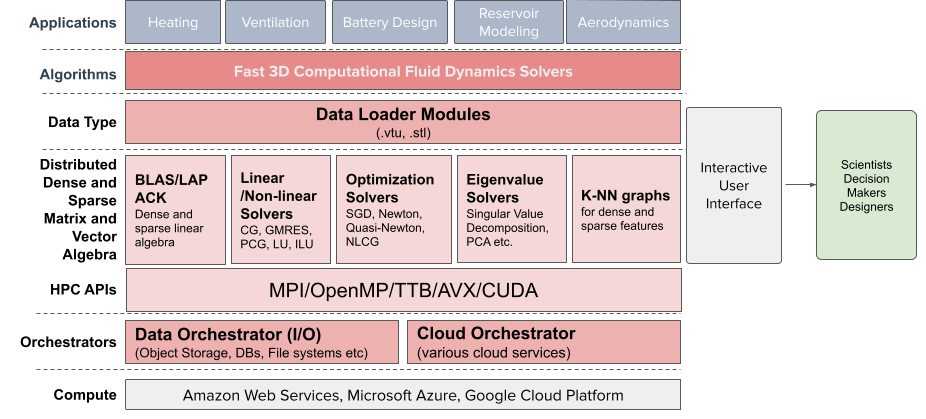}
    \caption{RocketML cloud orchestrator for large-scale CFD simulations.}
    \label{fig:rml_cloud_orchestrator}
\end{figure*}

Our target user for a cloud orchestrator is a researcher who has experience using an HPC cluster at a university or a national lab and has never used cloud computing. Ease-of-use and flexibility of using public cloud computing resources are the two most important requirements. Specifically, a cloud orchestrator that provides a graphical user interface (GUI) to 1) deploy and manage CPU- and GPU-based HPC clusters pre-installed with MPI, BLAS/LAPACK, and PETSc libraries along with a Slurm scheduler to simplify job management, 2) access to different IDEs like Jupyter Lab, VSCode for coding, compiling, and debugging HPC software, and 3) access to visualization software like Paraview for \emph{in situ} data visualization. We have explored several options available on Microsoft Azure and AWS marketplaces~\citep{azuremarketplace,awsmarketplace} and concluded that they either lack in ease-of-use or flexibility necessary for a researcher who has no exposure to cloud computing to conduct large-scale HPC experiments. In Figure \ref{fig:rml_cloud_orchestrator}, we show different components of RocketML’s cloud orchestrator that has been purpose-built to address a researcher’s needs to take complete advantage of HPC on a public cloud for scientific computing applications. 

\subsection{Mesh Creation}
\begin{figure*}
    \centering
    \includegraphics[width=1.0\textwidth]{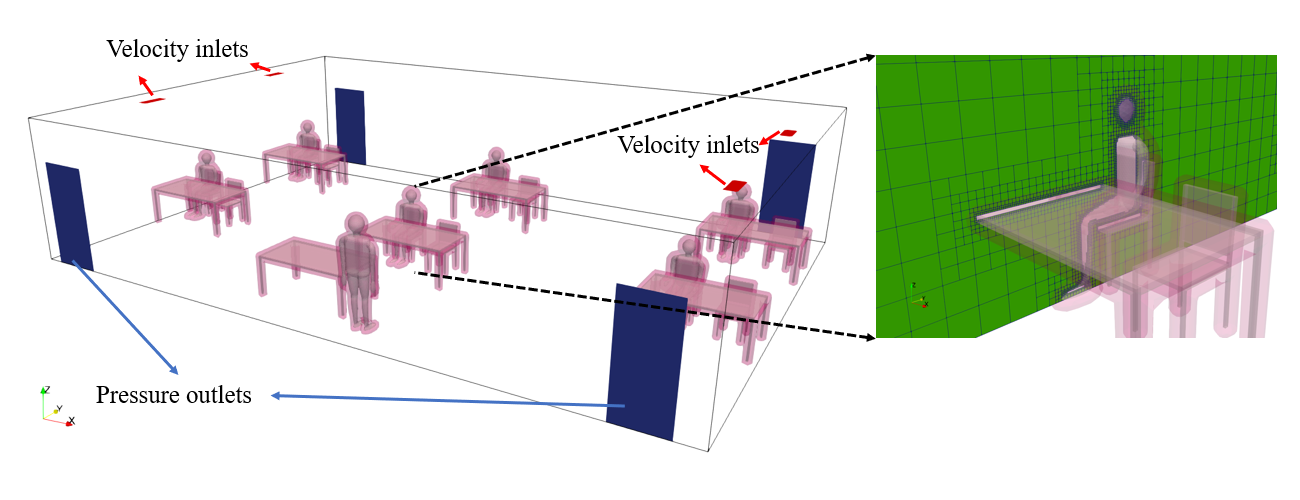}
    \caption{Computation domain and adaptive mesh generated from~\dkt{} framework.}
    \label{fig:domain}
\end{figure*}

We first begin with mesh creation. As mentioned in \secref{sec:background}, \dkt{} framework requires the specification of \In{} and \Out{} for a given queried point. The \In{}-\Out{} test is performed using Ray-Tracing. The input geometries are provided in the form of a \texttt{.stl} file format. Several instances of the same geometry, translated by some distance, can be grouped. An example of a config file for such an operation is:
\pagebreak

\begin{verbatim}
  geometries = (
  {
  mesh_path = "stl/human.stl"
  displacements = (
  {position = [0.65,0.0,0.8125]},
  {position = [1.625,0.0,0.73125]},
  )
  refine_lvl = 9
})
\end{verbatim}
where displacements are the shift from original position of \texttt{.stl} and \texttt{refine\_lvl} is the refinement level of the octree near the \texttt{.stl}. The octree generation starts with a coarse mesh and is subsequently refined until the required resolution is satisfied. We can similarly add multiple geometries. This approach gives an automated way to generate mesh for different scenarios. In the future, we plan to interface it with a GUI, where an application scientist can use the drag-and-drop feature to add multiple objects or to select the refine regions. \figref{fig:domain} shows the adaptive mesh generated for the classroom case.

\subsection{Solver}
We use \petsc{} as the linear solver for solving the system of equation AX=b. \petsc{} is a very well validated scalable linear algebra solver, which has a rich collection of preconditioners and solvers. Different \petsc{} options are configurable from the config file itself. An example is shown below:
\begin{verbatim}
solver_options_ht = {
  ksp_max_it = 500
  ksp_type = "bcgs"
  pc_type = "asm"
  ksp_atol = 1e-15
  ksp_rtol = 1e-15
  ksp_converged_reason = ""
}    
\end{verbatim}

\subsection{File I/O}
We use a parallel \texttt{.vtu} file format to write the data into the file at the user-defined interval. \texttt{.vtu} is an XML file format which enable a communication free write. Each processor writes its share of data in a separate file. All these files are later joined by another single \texttt{.pvtu} file, which is written by $0^{th}$ rank processor. 

\subsection{Checkpointing}
Machine failure is a common scenario when executing simulations on HPC servers. \dkt{} dumps the checkpoint file at user-defined intervals to keep a backup of the solutions. The checkpoint files are binary files that contain information about octree tree nodes and the solution vector. Similarly, to file I/O, each processor dumps its chunk of data in a communication - free fashion. If the code is killed for whatsoever reason, the user can restart the simulation by loading the checkpoint file.

Additionally, the user has the option to increase the number of nodes once the checkpoint is loaded. \dkt{} will re-partition the domain if the number of processors is increased. At this point, we have not implemented the ability to restart with a decreased number of processors.
\subsection{Visualization}
We have deployed \paraview{} within the cloud framework for in-situ visualization and related data--analysis. \paraview{} comes with \texttt{pvpython} (serial version) and  \texttt{pvbatch} (parallel version) which provide a seamless python interface.

As noted earlier, we rely on the config.txt file to pass input operation to the code. In future, we plan to replace it with a GUI interface hosted on the cloud platform for better interaction with the application user.

\section{Results}
\label{sec:Results}
\subsection{Roofline Analysis}
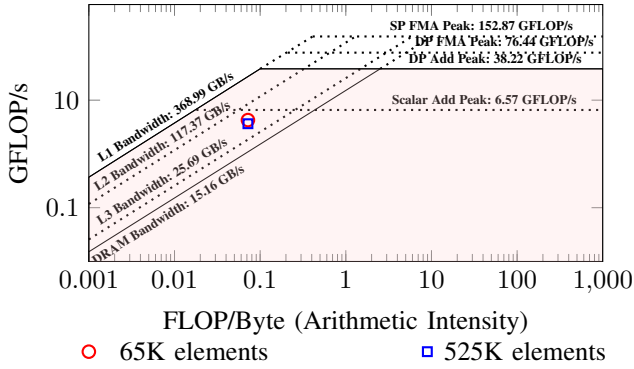
\begin{figure}
\centering
\begin{tikzpicture}
\begin{loglogaxis}[
    log ticks with fixed point,
    width=0.95\linewidth,height=5.0cm,
    xmax = 1000,
    xmin = 0.001,
    ymax = 600,
    ymin = 0.01,
    xlabel = FLOP/Byte (Arithmetic Intensity),
    ylabel = GFLOP/s
]

 \draw[-] (axis cs:0.001, 0.36899)  edge  node[sloped, anchor=left, above, text width=2.2cm, yshift=-0.05cm] {\fontsize{5pt}{4pt}\selectfont{\textbf{L1\;Bandwidth:\;368.99\;GB/s}}} (axis cs:0.1, 38.22);
\draw[dotted,thick] (axis cs:0.1, 38.22)  edge  node[sloped, anchor=left, above, text width=2.2cm] {} (axis cs:0.41, 152.87);
\draw[dotted,thick] (axis cs:0.001, 0.11737)  edge  node[sloped, anchor=left,xshift=-0.8cm, yshift=-0.05cm, above, text width=2.2cm] {\fontsize{5pt}{4pt}\selectfont{\textbf{L2\;Bandwidth:\;117.37\;GB/s}}} (axis cs:1.3, 152.87);

\draw[dotted,thick] (axis cs:0.001, 0.02569)  edge  node[sloped, anchor=left,xshift=-1.2cm, yshift=-0.05cm, above, text width=2.2cm] {\fontsize{5pt}{4pt}\selectfont{\textbf{L3\;Bandwidth:\;25.69\;GB/s}}} (axis cs:5.95, 152.87);

\draw[-] (axis cs:0.001, 0.01516)  edge  node[sloped, below,xshift=-1.2cm, yshift=0.05cm, text width=2.2cm] {\fontsize{5pt}{4pt}\selectfont{\textbf{DRAM\;Bandwidth:\;15.16\;GB/s}}} (axis cs:2.52, 38.22);

\draw[dotted,thick] (axis cs:2.52, 38.22)  edge  node[sloped, below, text width=2.2cm] {} (axis cs:10.08, 152.87);

\draw[dotted,thick] (axis cs:0.41, 152.87)  edge  node[sloped, above, text width=2.2cm,xshift=0.2cm, yshift=-0.05cm]{\fontsize{5pt}{4.2pt}\selectfont{\textbf{SP\;FMA\;Peak:\;152.87\;GFLOP/s}}} (axis cs:1000, 152.87);

\draw[dotted,thick] (axis cs:0.21, 76.44)  edge  node[sloped, above, text width=2.2cm,xshift=0.7cm,yshift=-0.05cm] {\fontsize{5pt}{4.2pt}\selectfont{\textbf{DP\;FMA\;Peak:\;76.44\;GFLOP/s}}} (axis cs:1000, 76.44);

\draw[-] (axis cs:0.1, 38.22)  edge  node[sloped, above, text width=2.2cm,xshift=0.8cm,yshift=-0.05cm] {\fontsize{5pt}{4.2pt}\selectfont{\textbf{DP\;Add\;Peak:\;38.22\;GFLOP/s}}} (axis cs:1000, 38.22);

\draw[dotted,thick] (axis cs:0.018, 6.57)  edge  node[sloped, above, text width=2.2cm,xshift=1.0cm,yshift=-0.05cm] {\fontsize{5pt}{4.2pt}\selectfont{\textbf{Scalar\;Add\;Peak:\;6.57\;GFLOP/s}}} (axis cs:1000,6.57);

\draw[fill=red!20,fill opacity=0.2] (axis cs:0.001, 0.001) -- (axis cs:0.001, 0.36899) --  (axis cs:0.1, 38.22) -- (axis cs:1000, 38.22) -- (axis cs:1000, 0.001) -- (axis cs:0.001, 0.001);

\node [circle,fill=none,fill opacity=0.4,draw,red, thick,minimum size=1pt,scale=0.5] at (axis cs:0.072,4.261) {};
\node [mark=square,fill=none,draw,blue,thick,minimum size=1pt,scale=0.5] at (axis cs:0.072,3.646){};

\end{loglogaxis}
\node [circle,fill=none,fill opacity=0.4,draw,red, thick,minimum size=1pt,scale=0.5] at (-0.,-1.2) {}; \node at (1.4,-1.2) {65K elements};

\node [mark=square,fill=none,draw,blue,thick,minimum size=1pt,scale=0.5] at (4.5,-1.2) {} ; \node at (5.8,-1.2) {525K elements};
\end{tikzpicture}
\caption{Figure showing roofline plot for the \mvec{} for linear basis function for two different meshes on \Frontera{}. The plot was generated using Intel Advisor.}
\label{fig:roofline}
\end{figure}
As the first step, we conducted a roofline analysis on \Frontera{} using Intel advisor to check whether the framework is compute-bound or memory-bound. Identifying this is important to identify and specify the type of cloud machines for deploying the framework. \figref{fig:roofline} shows the single-core roofline plot for the elemental \mvec{} computation using linear basis function on \Frontera{}. We limit ourselves to using linear basis functions as this mimics our application problem. We can see that the code is memory-bound, which is along the lines of what we expect for any finite element code. Overall, we can achieve a performance of about 4 GFLOP/s with an Arithmetic Intensity of 0.072, which corresponds to a bandwidth of approximately 60 GB/s. 

\begin{remark}
We have not used any hand-coded explicit vectorization to ensure the portability of the code across various platforms and have relied on compiler directives for vectorization.
\end{remark}

\subsection{Performance on the Cloud}

Based on the roofline analysis discussed in the previous section, we chose Azure virtual machines that are suited for memory-bound computational problems. This is an additional benefit of using cloud, where we can opt for a virtual machine with specific hardware configuration based on computational complexity of the problem at hand (memory-bound vs. compute-bound). Specifically, we chose the Azure Da and HBv2 series virtual machines that are built for memory-bound codes and compared the time-to-solution with the state-of-the-art HPC cluster \Frontera{}. \tabref{tab:azure_frontera_specs} provides the functional specifications of \Frontera{} and  virtual machines on cloud.

\begin{table}[!b]
\caption{Functional specifications of CPUs on Microsoft Azure and Frontera. \label{tab:azure_frontera_specs}}
\begin{center}
  \newcommand{\tabincell}[2]{\begin{tabular}{@{}#1@{}}#2\end{tabular}}
  \setlength\extrarowheight{3pt}
  \begin{tabular}{|c|c|c|c|}
    \hline
     \textbf{Specification} & \textbf{Da-series} & \textbf{HBv2-series}  & {\bf Frontera} \\
     \hline
     CPU & \tabincell{c}{AMD EPYC \\ 7452} & \tabincell{c}{AMD EPYC\\ 7V12} & \tabincell{c}{Intel Xeon \\ Platinum 8280} \\
     \hline
     Virtual CPU cores & 96 & 120 & 56 \\
     \hline
     Physical cores & 48 & 120 & 56 \\
     \hline
     Memory (GB) & 384 & 450 & 192\\
     \hline
     Clock rate (GHz) & 2.35 & 3.3 & 2.7 \\
     \hline
  \end{tabular}
\end{center}
\end{table}

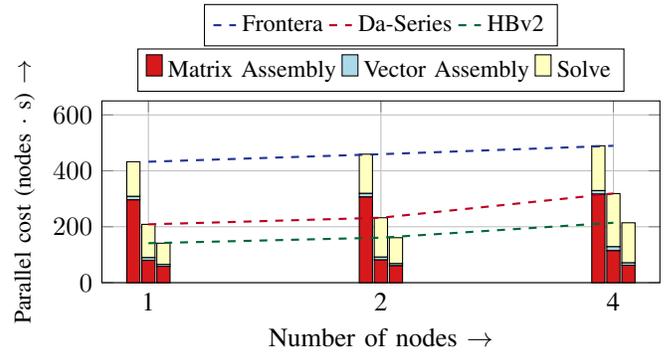
\begin{figure}[t!]
  \centering
  \begin{tikzpicture}
  \begin{axis}[
  ybar stacked, bar width=5pt,    
  xlabel={Number of nodes $\rightarrow$},
  ylabel={\small{Parallel cost (nodes $\cdot$ s) $\rightarrow$} },symbolic x coords={1,2,4},width=9cm,height=4cm,
  xtick = data, 
  ymin=0,
  ymax=650,
  legend pos=north west,grid=major,legend style={at={(0.5,1.05)},
		anchor=south,legend columns=5}]

  \addplot [fill=div_d1] [bar shift=-0cm] table[x={Nodes} , y expr={\thisrow{MatAssembly}*\thisrow{Nodes}} ]{Data/NSHT/DSeries.txt};
  \addplot [fill=div_d4] [bar shift=-0cm]  table[x={Nodes}, y expr={\thisrow{VecAssembly}*\thisrow{Nodes}}]{Data/NSHT/DSeries.txt};
  \addplot [fill=div_d3] [bar shift=-0cm]  table[x={Nodes}, y expr={(\thisrow{Solve} - \thisrow{MatAssembly} - \thisrow{VecAssembly}) *\thisrow{Nodes}}]{Data/NSHT/DSeries.txt};
  
  \makeatletter
  \newcommand\resetstackedplotsSix{
  \makeatletter
  \pgfplots@stacked@isfirstplottrue
  \makeatother
  \addplot [forget plot,draw=none] coordinates{(1,0) (2,0) (4,0)};
 }
  \makeatother
  \resetstackedplotsSix
\addplot [fill=div_d1] [bar shift=0.2cm] table[x={Nodes} , y expr={\thisrow{MatAssembly}*\thisrow{Nodes}} ]{Data/NSHT/HSeries.txt};
  \addplot [fill=div_d4] [bar shift=0.2cm]  table[x={Nodes}, y expr={\thisrow{VecAssembly}*\thisrow{Nodes}}]{Data/NSHT/HSeries.txt};
  \addplot [fill=div_d3] [bar shift=0.2cm]  table[x={Nodes}, y expr={(\thisrow{Solve} - \thisrow{MatAssembly} - \thisrow{VecAssembly})*\thisrow{Nodes}}]{Data/NSHT/HSeries.txt};
  
  \legend{\small{Matrix Assembly}, \small{Vector Assembly}, \small{Solve}}
  
    \makeatletter
  \newcommand\resetstackedplotsSeven{
  \makeatletter
  \pgfplots@stacked@isfirstplottrue
  \makeatother
  \addplot [forget plot,draw=none] coordinates{(1,0) (2,0) (4,0)};
 }
  \makeatother
  \resetstackedplotsSix
\addplot [fill=div_d1] [bar shift=-.2cm] table[x={Nodes} , y expr={\thisrow{MatAssembly}*\thisrow{Nodes}} ]{Data/NSHT/Frontera.txt};
  \addplot [fill=div_d4] [bar shift=-.2cm]  table[x={Nodes}, y expr={\thisrow{VecAssembly}*\thisrow{Nodes}}]{Data/NSHT/Frontera.txt};
  \addplot [fill=div_d3] [bar shift=-.2cm]  table[x={Nodes}, y expr={(\thisrow{Solve} - \thisrow{MatAssembly} - \thisrow{VecAssembly})*\thisrow{Nodes}}]{Data/NSHT/Frontera.txt};
  \end{axis}

    \begin{axis}[ axis y line=none,axis x line = none,symbolic x coords={1,2,4},width=9cm,height=4cm,xtick = data,legend pos={north west},legend style={at={(0.0,1.65)}},legend columns=4,
    ymin=0,
  ymax=650,
  legend pos=north west,grid=major,legend style={at={(0.5,1.3)},
		anchor=south,legend columns=5}
  ] 
      \addplot[sq_b1,thick,dashed]  table[x ={Nodes}, y expr={\thisrow{Solve}*\thisrow{Nodes}} ]{Data/NSHT/Frontera.txt};
      \addplot[sq_r1,thick,dashed]  table[x={Nodes}, y expr={\thisrow{Solve}*\thisrow{Nodes}} ]{Data/NSHT/DSeries.txt};
      \addplot[sq_g1,thick,dashed]  table[x={Nodes}, y expr={\thisrow{Solve}*\thisrow{Nodes}} ]{Data/NSHT/HSeries.txt};
      
    
      \legend{\small{Frontera},\small{Da-Series},\small{HBv2}}
  \end{axis}  
  
  \end{tikzpicture}
  \caption{\textit{Strong scaling for Navier Stokes  + Heat Transfer:\;} Figure showing the parallel cost (time $\times$ nodes) variation with increase in the number of nodes. A straight line parallel to X-axis indicates ideal scaling.}
  \label{fig:strong-scaling}
\end{figure}

\textbf{Scaling behavior comparison:} To study the scaling behavior and time comparison of our solver on these machines, we consider the classroom case (shown in Fig.~\ref{fig:domain}) and solve the Navier--Stokes equation coupled with Heat Transfer. In both the azure machines (Da series and HBV2 series), we only used 50\% of the available CPU cores per node (48 for Da series and 60 for HBv2 series) mentioned in the \tabref{tab:azure_frontera_specs}. We note that the specs take the hyper-threading into account for Da-series with 2 threads per core. Since our code is only MPI parallelized, we are able to use only half of the available cores. On \Frontera{} we used all 56 cores available per node. The overall mesh for scaling analysis consisted of around 790 K elements with higher refinement near the body (mannequin/tables) and inlet regions.  We compared the three main components of the solver: matrix assembly, vector assembly, and the total solve time. \figref{fig:strong-scaling} shows the parallel cost (defined as total time multiplied by the number of nodes used) and time taken on these three different machines. A straight line parallel to X-axis indicates ideal scaling. We can see that the cloud machines take significantly less time than \Frontera{}, with the HBv2 series taking the least amount of time. In particular, 10 time steps of the solver took 432s on 1 node of \Frontera{}, while it took 132s on 1 node of HBv2 series machine.

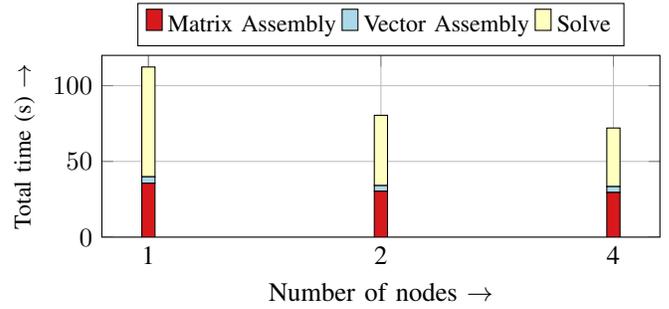
\begin{figure}[t!]
  \centering
  \begin{tikzpicture}
  \begin{axis}[
  ybar stacked, bar width=5pt,    
  xlabel={Number of nodes $\rightarrow$},
  ylabel={\small{Total time (s) $\rightarrow$} },symbolic x coords={1,2,4},
  width=9cm,height=4cm,
  xtick = data, 
  ymin=0,
  ymax=120,
  legend pos=north west,grid=major,legend style={at={(0.5,1.05)},
		anchor=south,legend columns=5}]

  \addplot [fill=div_d1] [bar shift=-0cm] table[x={Nodes} , y expr={\thisrow{MatAssembly}} ]{Data/NSHT/MemoryBandwidth.txt};
  \addplot [fill=div_d4] [bar shift=-0cm]  table[x={Nodes}, y expr={\thisrow{VecAssembly}}]{Data/NSHT/MemoryBandwidth.txt};
  \addplot [fill=div_d3] [bar shift=-0cm]  table[x={Nodes}, y expr={(\thisrow{Solve} - \thisrow{MatAssembly} - \thisrow{VecAssembly})}]{Data/NSHT/MemoryBandwidth.txt};
   \legend{\small{Matrix Assembly}, \small{Vector Assembly}, \small{Solve}}
 \end{axis}
  
  \end{tikzpicture}
  \caption{\textit{Effect of memory bandwidth\;} Time comparison for different distribution of processor across nodes. Total number of processor was kept constant to 120, whereas the number of processor per node was varied.}
  \label{fig:memBandwidth}
\end{figure}
\begin{figure*}[t!]

\begin{subfigure}{0.33\textwidth}
\centering
\includegraphics[width=\linewidth]{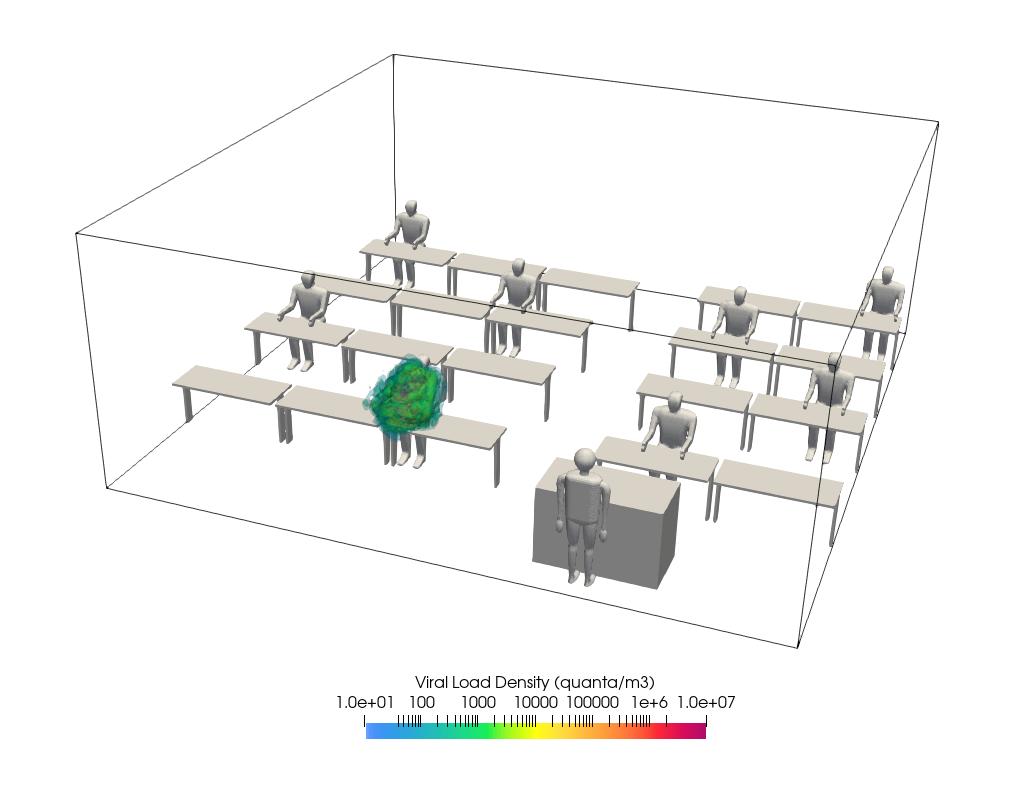}
\vspace{-0.3in}
\caption{t=2s}
\label{fig: mannequin2_2s}
\end{subfigure}
\begin{subfigure}{0.33\textwidth}
\centering
\includegraphics[width=\linewidth]{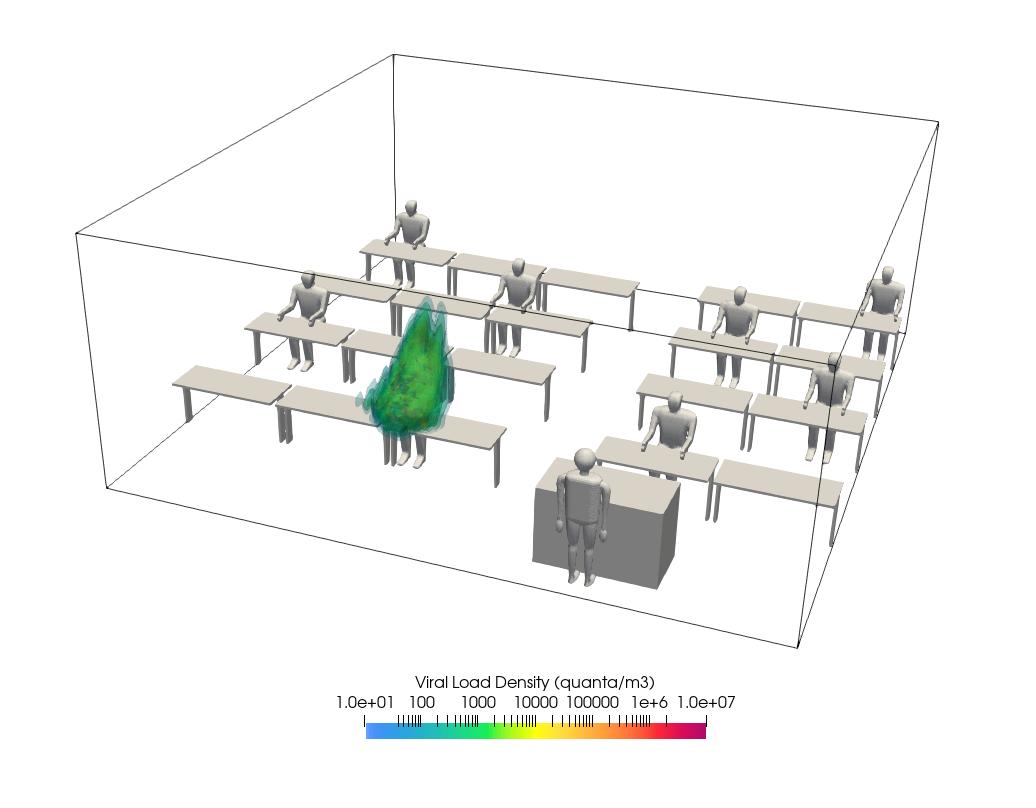}
\vspace{-0.3in}
\caption{t=5s}
\label{fig: mannequin2_5s}
\end{subfigure}
\begin{subfigure}{0.32\textwidth}
\centering
\includegraphics[width=\linewidth]{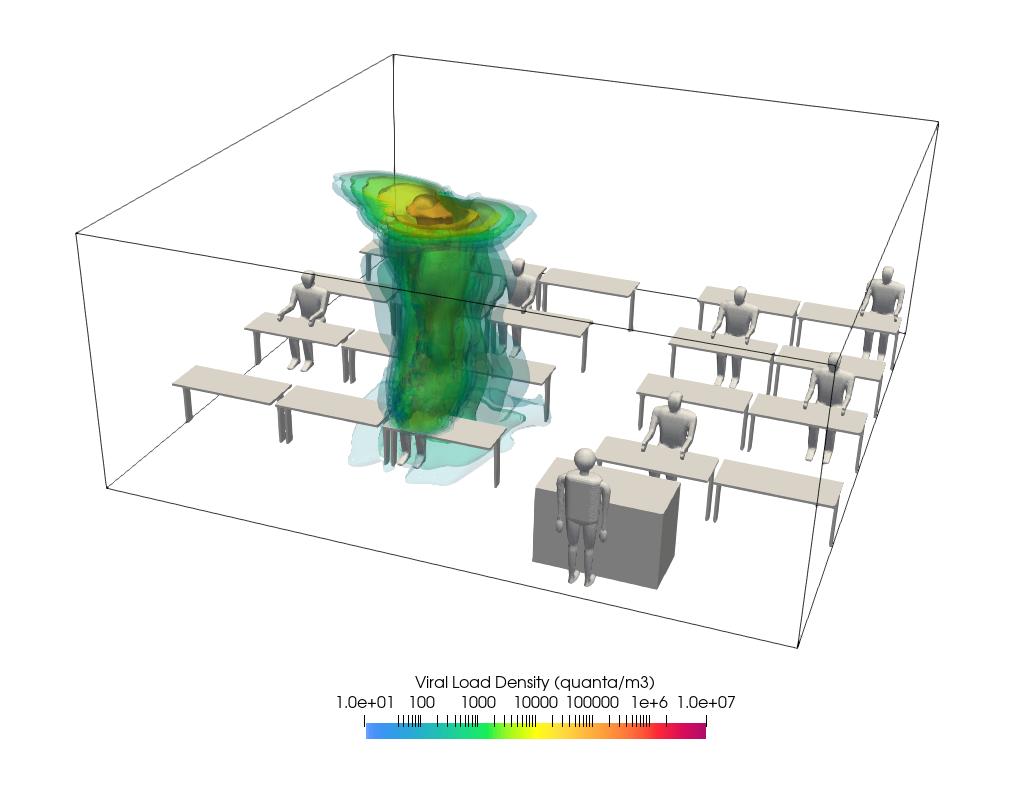}
\vspace{-0.3in}
\caption{t=25s}
\label{fig: mannequin2_10s}
\end{subfigure}
\vskip\baselineskip 
\begin{subfigure}{0.33\textwidth}
\centering
\includegraphics[width=\linewidth]{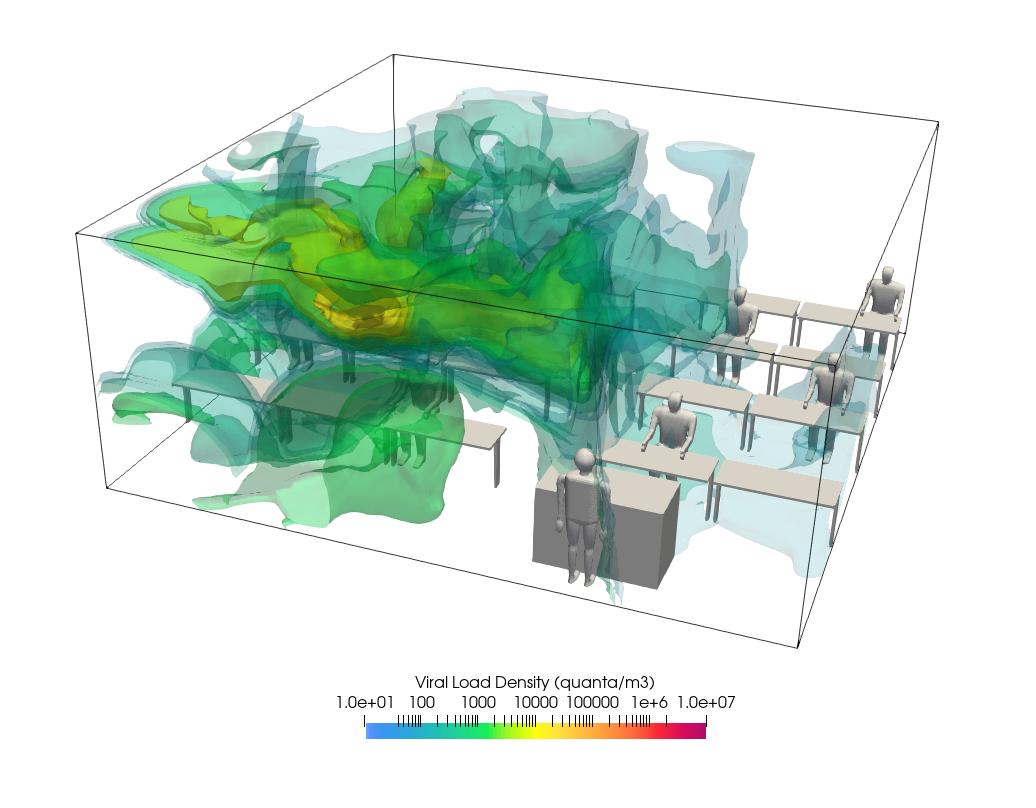}
\vspace{-0.3in}
\caption{t=150s}
\label{fig: mannequin2_25s}
\end{subfigure}
\begin{subfigure}{0.33\textwidth}
\centering
\includegraphics[width=\linewidth]{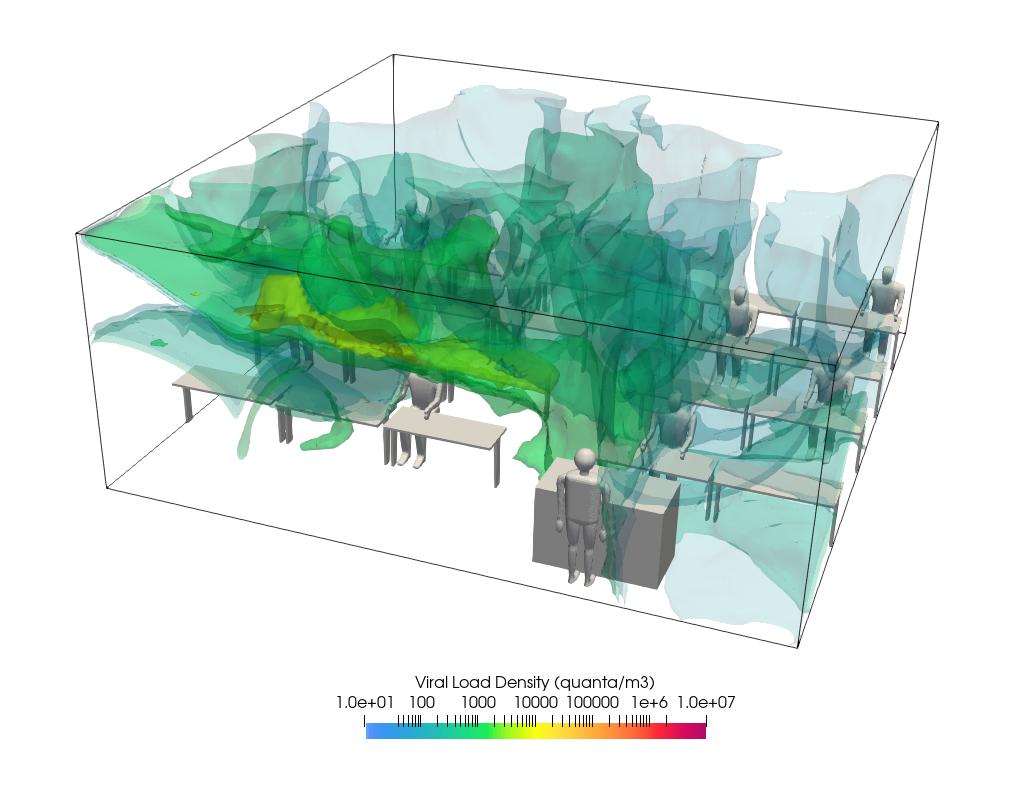}
\vspace{-0.3in}
\caption{t=300s}
\label{fig: mannequin2_50s}
\end{subfigure}
\begin{subfigure}{0.33\textwidth}
\centering
\includegraphics[width=\linewidth]{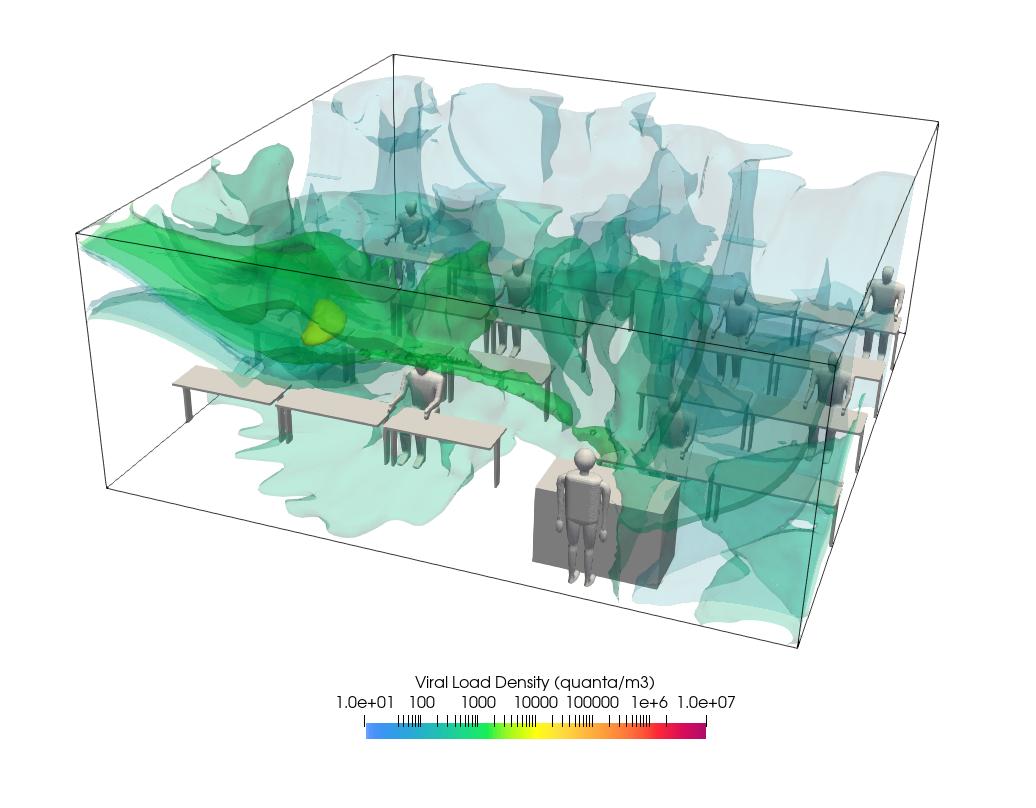}
\vspace{-0.3in}
\caption{t=450s}
\label{fig: mannequin2_100s}
\end{subfigure}
\vspace{0.1in}
\caption{Time evolution of the viral load concentration of the air due to a simulated cough by a mannequin.}
\label{fig: concentrationTime_mannequin2}
\end{figure*}

\textbf{Explanation of observed scaling behavior:} The two main architectural factors responsible for the difference between Azure machines and \Frontera{} are memory bandwidth and cache size. The total peak memory bandwidth on the HBv2 node is about 340GB/s and around 210GB/s on a Frontera CLX node. For memory-bound workloads like CFD simulations, the higher memory bandwidth on the HBv2 node would result in a better solve time. Each AMD EPYC 7742 CPU in HBv2 VMs has 256MB of L3 cache divided across 16 core complexes with four cores each (4MB L3 per core, 4.5MB L2+L3 per core). Conversely, the Intel 8280 has only 38.5MB of L3 cache (1.4MB L3 per core, 2.4MB L2+L3 per core). Codes that benefit from high cache reuse would enormously benefit from the AMD Zen2 architecture, resulting in a much higher “effective” memory bandwidth. This indicates that an educated choice of the architecture for the code can lead to a significant boost in the performance. Such flexibility is offered only by cloud, and not on traditional HPC based clusters.

\begin{remark}
We note that our code is general-purpose and does not contain any explicit vectorization cache blocking or streaming operation, ensuring code portability. This approach is common in most of the FEM / fluid dynamics code. Explicit optimization such as vectorization/cache blocking can significantly boost the time to solve but comes at the cost of portability on different machines. Comparing the performance of hardware-specific optimized codes on these machines is beyond the scope of the work.
\end{remark}

\textbf{Parallel efficiency:} We also see a good parallel efficiency on all these machines. Overall \Frontera{} provides the best parallel efficiency (0.88) with the solve time decreasing from 432s on 1 node to 122s on 4 nodes. The cloud machines have a relatively low parallel efficiency of about 0.65. The deviation from ideal scaling can be attributed to the increased iteration count for the preconditioned solve. Interested readers are referred to~\citep{saurabh2020industrial} for detailed discussion. Briefly, it is expected that bare-metal HPC servers will have better parallel efficiency than virtual-machine-based cloud HPC servers due to increased communication costs across nodes. Optimizing multi-node VM-based HPC servers suggests the need to deploy new algorithms that minimizes communication at the cost of increased (and/or redundant) computations. 

\textbf{Memory bandwidth effect:} Memory bandwidth plays a pivotal role in overall solve time, especially for memory bound codes. Therefore, we compare the effect of memory bandwidth on our solver. The number of processors is kept constant at 120, whereas the number of nodes is varied from 1 to 4 with a varying number of processors per node. Note that 4 nodes have 30 processors per node, and 1 node has 120 processors. With the increase in the number of processors per node, the memory bandwidth per core decreases. For instance, on the HB120rs\_v2 (HBV series)  machine, the memory bandwidth per core decreases from 11.50 GB/s when using 30 cores/node to 2.88 GB/s while using all 120 cores.~\citep{AzureHBV2}. \figref{fig:memBandwidth} shows the effect of the memory bandwidth of our solver.  We see a strong correlation between memory bandwidth and solve time. With the increase in the memory bandwidth, the solve time decreases from around 112 s with 120 cores/node to 80 s on 30 cores/node. However, we note that this reduction in time comes at an increased cost of utilizing more nodes. 

\section{Application Results}
\label{sec:ApplicationResult}
\begin{figure*}[t!]
\centering
\begin{subfigure}{.45\textwidth}
  \centering
  \includegraphics[width=1.0\linewidth,trim=100 0 80 200, clip]{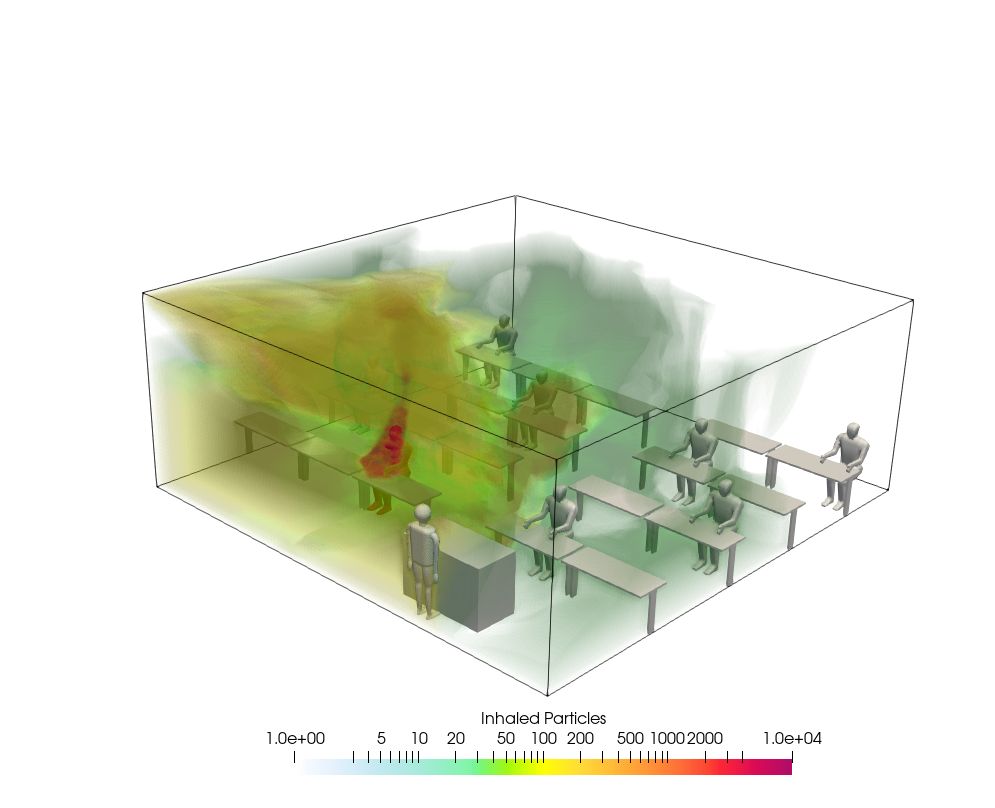}
  \caption{Without computer monitors}
  \label{fig:classroom:noMonitors}
\end{subfigure}%
\begin{subfigure}{.45\textwidth}
  \centering
  \includegraphics[width=1.0\linewidth,trim=100 0 80 200, clip]{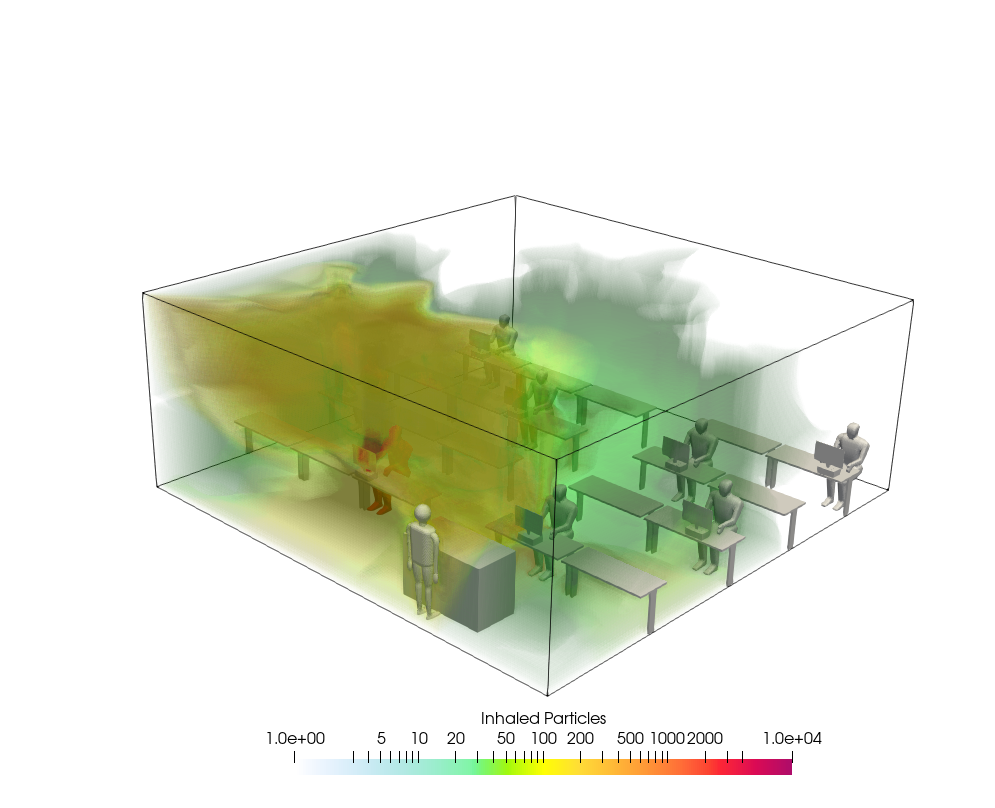}
  \caption{With computer monitors}
  \label{fig:classroom:Monitors}
\end{subfigure}
\par\bigskip
\begin{subfigure}{.45\textwidth}
  \centering
  \includegraphics[width=1.0\linewidth,trim=100 0 80 200, clip]{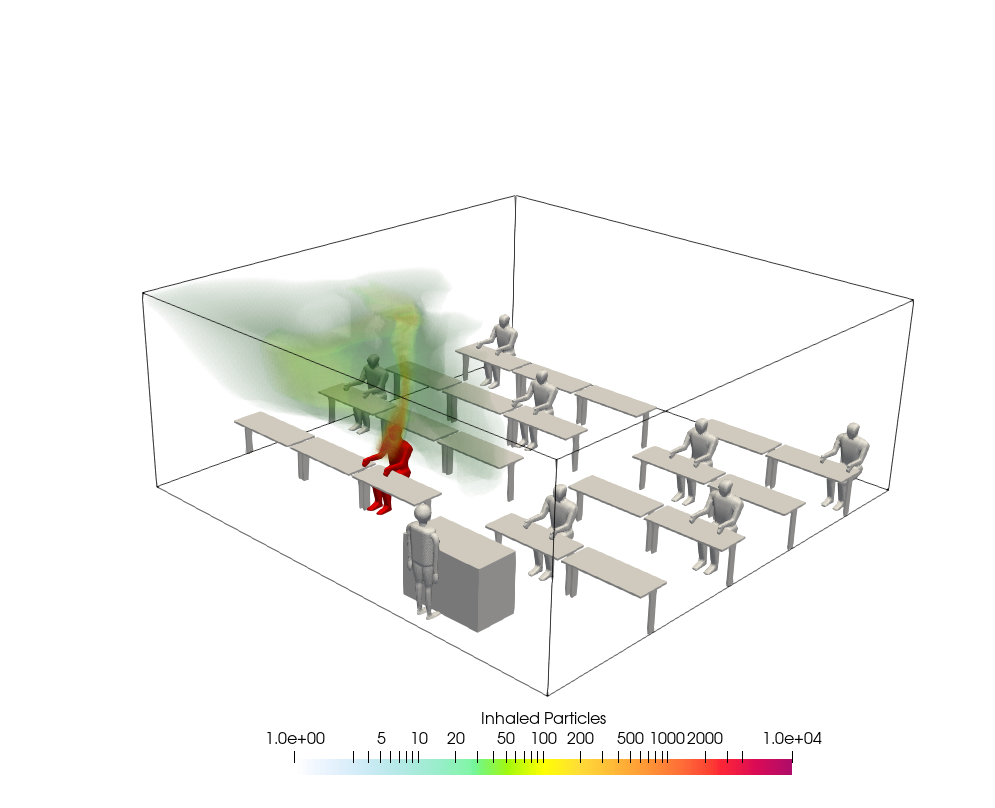}
  \caption{Everyone wearing masks}
  \label{fig:classroom:Masks}
\end{subfigure}
\begin{subfigure}{.45\textwidth}
  \centering
  \includegraphics[width=1.0\linewidth,trim=100 0 80 200, clip]{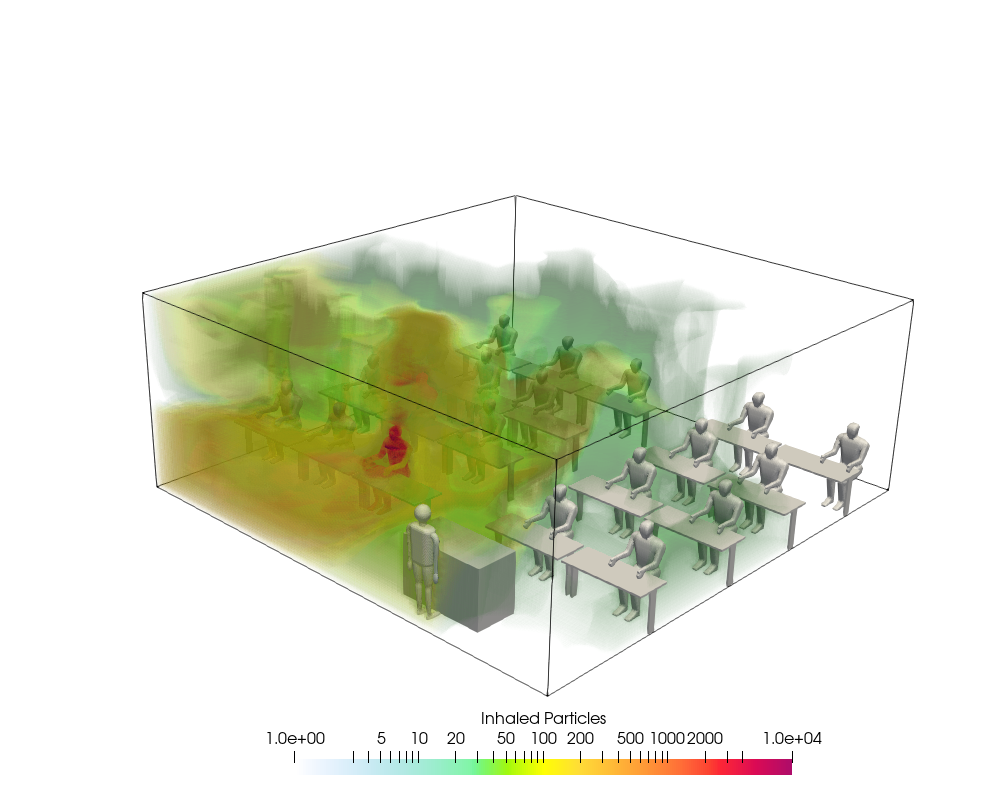}
  \caption{Increased occupancy}
  \label{fig:classroom:increasedOccupancy}
\end{subfigure}
\vspace{2 mm}
\caption{\textit{Classroom scenario:} Evaluation of inhaled particles in different classroom scenarios. The mannequin marked in the red is infected with COVID and transmits the virus. The isocontours represent the regions of different inhaled particle concentrations in space.}
\label{fig:classroom}
\end{figure*}

In this section, we illustrate results of our application case. We consider a typical university classroom as illustrated in  Fig.~\ref{fig:domain}. The room has a length of 9~${m}$, a width of 9~${m}$, and a height of 3.5~${m}$ with a ventilation system  installed on the ceiling. The occupants are modeled using mannequins. A representative resolved mesh is shown as an inset in this figure. The Reynolds number based on the inlet air velocity and classroom height is estimated to be $10^5$. The spatial discretization was performed using linear basis function and the time discretization was performed using second order BDF2. A dimensionless timestep of 1E-3 was used to carry out the simulation. Each scenerio takes up to a final time of 2 dimensionless time units to reach the stationary steady state (2K timestep), which required about a day of compute time on HBV2 series machine. This resulted in the total cost of \$ 340 for simulating each scenario.

We evaluate the impact of one infected individual who coughs, releasing an aerosolized load of viral particles. \figref{fig: concentrationTime_mannequin2} plots the time evolution of the viral load concentration of the air due to a simulated cough by a mannequin. Notice that the aerosolized concentration plume rises and is recirculated by the inlet air. The mixing with the inlet air dilutes the virus concentration, but the inlet air helps spread the concentration across the room, potentially impacting everyone in the classroom.

 Finally, \figref{fig:classroom} shows the isocontours of viral particles concentration (at one time-point) for different operating scenarios. We compare the different scenarios in the classroom with (\figref{fig:classroom:noMonitors}) and without (\figref{fig:classroom:Monitors}) monitors, increased occupancy (\figref{fig:classroom:increasedOccupancy}) and the situation where everyone wears a mask (\figref{fig:classroom:Masks}). The mannequin colored in red is the infected person. The regions of contours with inhaled particles $>$ 50 ~\citep{kolinski2021superspreading} are considered at a high risk of transmission. We can see different scenarios lead to significantly different regions of risk. Additionally, the wearing of masks significantly reduced the risk of transmission and should therefore be encouraged. We defer a detailed analysis and implication of these results to a companion publication.


\section{Conclusions}
In this work, we deploy a scalable, adaptive framework on the cloud utilizing \dkt{} as a mesh generation tool and \petsc{} as a scalable linear algebra solver to analyze the risk of COVID-19 transmission in different classroom scenarios. The problem utilizes running different simulation instances, which are easily deployable in a cloud-based setting. Further, we analyzed the performance of the cloud-based resources to the state-of-the-art supercomputer, specifically \Frontera{}. With the proper choice of cloud machines, we observe on par performance on the cloud, if not better, with the supercomputers. This suggests that cloud based VM HPC servers offer a viable approach to democratize and deploy complex simulation workflows. In the future, we would like to simplify the deployment of the use-cases with more GUI-friendly interfaces.

\section{Acknowledgments}
We acknowledge partial support from NSF 1855902.












\bibliographystyle{IEEEtran}
\bibliography{./main}


\end{document}